\documentclass[aps,prd,tightenlines,preprint,nofootinbib,a4paper]{revtex4}
\usepackage{epsfig}

\begin{document}

\title{Flavor mixing and the permutation symmetry among generations}

\author{T. K. Kuo\footnote{tkkuo@purdue.edu}}
\affiliation{Department of Physics, Purdue University, West Lafayette, IN 47907, USA}

\author{S. H. Chiu\footnote{schiu@mail.cgu.edu.tw}}
\affiliation{Physics Group, CGE, Chang Gung University, 
Taoyuan 33302, Taiwan}

\begin{abstract}

In the standard model, the permutation symmetry among the three generations of fundamental fermions is
usually regarded to be broken by the Higgs couplings. It is found that the symmetry is restored if we
include the mass matrix parameters as physical variables which transform appropriately under  
the symmetry operation.  Known relations between these variables, such as the renormalization group
equations, as well as formulas for neutrino oscillations (in vacuum and in matter), are shown to be
covariant tensor equations under the permutation symmetry group.

\end{abstract}


\maketitle

\pagenumbering{arabic}



\section{Introduction}

One of the long-standing puzzles in the Standard Model (SM) is the existence of three
generations of fermions which behave identically under the gauge interactions.
The resulting exchange symmetry will be dubbed as the g-permutation symmetry in this paper.
This symmetry is commonly regarded as broken by the Higgs coupling through its vacuum expectation
value (VEV), resulting in four mass matrices and a plethora of physical parameters, viz., 
the fermion masses and the mixing matrices for quarks ($V_{CKM}$) and for neutrinos ($V_{PMNS}$).
If these parameters are considered as fixed entities, then they would seem to be a collection of arbitrary
numbers, which do not transform under the symmetry operation,
and the g-permutation would just be broken. However, there are at least three classes of
physical phenomena which suggest an alternative interpretation. 1) Neutrino oscillation in vacuum.
Here, as a neutrino beam travels, the mixing parameters evolve along and are not static. 2) Neutrino oscillation
in matter (see, e.g., \cite{Wolfenstein:1977ue,Mikheev:1986gs}).  
When neutrinos propagate in a medium, an induced mass is obtained which changes the mixing pattern.
3) The renormalization group equations (RGE) for quarks 
(see, e.g., \cite{Machacek:1983fi,Sasaki:1986jv,Babu:1987im,Chiu:2008ye,Chiu:2016qra})   
and for neutrinos (see, e.g., \cite{Lindner:2005as,Chiu:2015ega}).   
A change in energy scales entails a new set of parameters and are governed by the RGE.  For cases 2) and 3),
one could say that the physical ``vacuum" itself is evolving.  In all of these examples, conceptually, it is
more natural to regard the mass matrix parameters as changeable physical variables.  And, when one considers
g-permutation, these variables should also transform under the symmetry operations.  Once we do that,
it becomes clear that they have natural assignments as tensors under $S_{3}$, the permutation group of three objects.
With this interpretation one can show that the g-permutation symmetry, now operating on both the fundamental fermions and the mass matrix parameters, is restored.
In this connection it should be noted that, in SM, the mass parameters are, 
apart from a common Higgs VEV, identified with the ``coupling constants" of the Higgs to the fermions.
The permutation operation, $(\psi_{i},m_{i}) \rightarrow (\psi_{j},m_{j})$, is not unlike the charge
conjugation transformation, $(\psi,e) \rightarrow (\psi^{c},-e)$.
Thus it is reasonable to include masses in a permutation operation.

In the literature, there are numerous relations amongst the mass matrix parameters associated with
neutrino oscillations and RGE of quarks and neutrinos 
(see, e.g., \cite{Chiu:2017ckv,Chiu:2016qra,Chiu:2015ega}, and the references therein).   
These are obtained by direct and explicit calculations.
When written in appropriate variables, hints of a permutation symmetry seem ubiquitous.
In this paper we present a general analysis of these equations.  
It is found that the SM has a g-permutation symmetry group $S_{3}(u) \times S_{3}(d) \times S_{3}(l) \times S_{3}(\nu)$
(or $[S_{3}]^{4}$), where the factors denote permutations in sectors of the $u-$ and $d-$type quarks,
the charged leptons and the neutrinos, respectively.  Also, the relations mentioned above are all covariant tensor equations
under $[S_{3}]^{4}$, just like the tensor equations in theories with rotational symmetry.
It should be emphasized that, to establish a symmetry for a given Lagrangian, it is
necessary to assign appropriate transformation properties to the variables contain therein.
The symmetry $[S_{3}]^{4}$ would be broken if one assigns the mass matrix variables as singlets under its operation.

Our considerations are similar to those of another example of symmetry restoration in a familiar setting.
Consider the case of an atom in an external $\vec{B}$-field.  The interaction term is proportional to
$\vec{B} \cdot \vec{\sigma}$.  If we treat $\vec{B}$ as a fixed external field, then this term breaks the rotational symmetry.
On the other hand, we can include $\vec{B}$ as a dynamical variable in the atomic system, transforming as a vector,
then rotational symmetry is restored. The transformation of the mass matrix parameters under g-permutation
is analogous to the rotation of the $\vec{B}$-field.

We add that the Particle Data Group (PDG) parametrization \cite{PDG} is ill-equipped to exploit the g-permutation symmetry.   
Besides being rephasing dependent, the PDG variables $\theta_{ij}$'s, despite their appearances, have
very complicated behaviour under g-permutation, making it difficult to uncover possible symmetries in an equation.

\section{Tensor Analysis of $S_{3}$}

We turn now to an analysis of the representation of the g-permutation group, which is based on $S_{3}$.
The elements of $S_{3}$ operate on three objects, say $B_{i}$, $i=1,2,3$.  For our purposes, it suffices to concentrate
on the exchange operators: 
\begin{equation}
X_{ij}: B_{i}\leftrightarrow B_{j}, B_{k}\leftrightarrow B_{k},  i\neq j \neq k.
\end{equation}
To borrow the terminology of $O(3)$, we will call $B_{i}$ a P-vector or $B_{i} \sim \bf{3}$.
The three-dimensional representation of $S_{3}$, however, is reducible ($\sum B_{i}=$ invariant). 
Nevertheless, it is convenient to use the reducible \textbf{3} and develop a tensor analysis for $S_{3}$,
similar to that for $O(3)$.  This is useful because, as it turns out, the physical variables behave
like P-tensors under permutations, and relations between them are covariant P-tensor equations.

To begin, we note that, different from the linear algebra of $O(3)$,
simple functions of $B_{i}$ behave like $B_{i}$ under permutations, and are also P-vectors.  E.g.,
\begin{eqnarray}
(1/B_{i}) &= &(1/B_{1},1/B_{2},1/B_{3}) \sim \bf{3}  \nonumber \\
(\sin B_{i}) &=& (\sin B_{1},\sin B_{2}, \sin B_{3}) \sim \bf{3} \nonumber \\
(\sin^{2} B_{i})&=&(\sin^{2} B_{1}, \sin^{2}B_{2},\sin^{2}B_{3}) \sim \textbf{3}, \hspace{0.1in} etc.
\end{eqnarray}

Next, out of two P-vectors, $B_{i}$ and $C_{i}$, we can construct rank-two P-tensors such as
$f(B_{i}) \pm f(C_{j})$ or $f(B_{i})f(C_{j})$, where $f$ is some regular function.
The simplest of these tensors are $B_{i}\pm C_{i}$ or $B_{i}C_{j}$.
Thus, the product $B_{i}C_{j}$ can be decomposed into three P-vectors: 
1) Diagonal: $(B_{1}C_{1},B_{2}C_{2},B_{3}C_{3})=D_{ii}$ (no sum); 
2) Symmetrical: $S_{ij}=B_{i}C_{j}+B_{j}C_{i}=S_{ji}, i\neq j$; 
3) Anti-symmetrical: $A_{ij}=B_{i}C_{j}-B_{j}C_{i}=-A_{ji}, i\neq j$.
Their transformation properties may be further elucidated by the use of invariant tensors in $S_{3}$.
In addition to the familiar $O(3)$ tensors $\delta_{ij}$ and $e_{ijk}$, in $S_{3}$ there is a third,
$E_{ijk}$ which is defined as: 
\begin{eqnarray}
E_{ijk}=\left \{\begin{array}{ll}
1, & i\neq j \neq k, \mbox{even under exchange of indices},   \\
0, & \mbox{any repeated index}.
\end{array}
\right.
\end{eqnarray}
This may be dubbed as the ``symmetrical Levi-Civita symbol".
Using these, we have the following:
\begin{eqnarray}
\delta_{ij}B_{i}C_{j}&=&\sum_{i} D_{ii} \sim \bf{1}, \hspace{0.1in} \mbox{P-scalar}; \nonumber \\
E_{ijk}S_{jk} &\sim& \bar{S}_{i} \sim \textbf{3}, \hspace{0.1in} \mbox{P-vector}; \nonumber \\
e_{ijk}A_{jk} &\sim& \widetilde{A}_{i} \sim  \widetilde{\textbf{3}}, \hspace{0.1in} \mbox{pseudo-P-vector}.
\end{eqnarray}
$\widetilde{A}_{i}$ is a pseudo-P-vector since under the exchange operator,
\begin{equation}
X_{ij}:\widetilde{A}_{i} \leftrightarrow -\widetilde{A}_{j}, \widetilde{A}_{k} \leftrightarrow -\widetilde{A}_{k}.
\end{equation}
Thus, the rank-two P-tensor $B_{i}C_{j}$ is decomposed into three \textbf{3}'s under $S_{3}$,
two of them are P-vectors, while the third is a pseudo-P-vector.

Other useful constructions are 
\begin{equation}
F=E_{ijk}B_{i}C_{j}D_{k} \sim \textbf{1}, \hspace{0.1in} \mbox{P-scalar} \hspace{0.1in}  (X_{ij}:F \rightarrow +F),
 \end{equation}
 \begin{equation}
 G=e_{ijk}B_{i}C_{j}D_{k} \sim \widetilde{\textbf{1}}, \hspace{0.1in} \mbox{pseudo-P-scalar} \hspace{0.1in} (X_{ij}:G \rightarrow -G).
\end{equation}
In addition, for odd or even functions of $A_{ij}$, e.g., 
\begin{equation}
   \sin A_{ij} \sim \widetilde{\textbf{3}}, \hspace{0.1in}  \sin^{2} A_{ij} \sim \textbf{3}.  
 \end{equation}
 
In summary, the tensor analysis of $S_{3}$ has a lot in common with that of $O(3)$,
though with two important differences: 1) The existence of the symmetric Levi-Civita
symbol $E_{ijk}$; 2) The linear tensor algebra is generalized to include functions of tensors for $S_{3}$.
Once these two differences are properly managed, the implementation of the $S_{3}$ symmetry amounts
to demanding that all relations are covariant tensor equations, just like the familiar equations which
are covariant under rotation.  
   
\section{The Broken g-Permutation Symmetry and Its Restoration}

We will now turn to the g-permutation symmetry in the SM.
To accommodate neutrino oscillations which are central to the consideration in this paper, the SM will be
augmented by the inclusion of Dirac neutrino mass term. This minimal extension of the SM brings the leptonic sector 
on a par with the quark sector, and will facilitate the ensuing discussions.  The interesting possibility of neutrinos
being Majorana particles will not be dealt with here, but could hopefully be the topic of a future investigation. 

In order not to clutter our notations, we will first concentrate our discussion to the leptonic sector of the SM.
The parallel case of the quark sector will be brought in when appropriate. 
Also, to study the effect of g-permutation, one need only to focus on the part of the SM Lagrangian which
contains the fermion-Higgs interaction, after it has acquired its VEV.  The result, after diagonalization of the mass matrix,
can be represented by the following terms in the Lagrangian (see, e.g., Ref. \cite{Ramond}), schematically,   
\begin{eqnarray}
\mathcal{L}(l,H) &\sim& (J_{\mu} W^{\dag}_{\mu} + h.c.)-(1+\frac{h}{v})
(\sum_{\alpha}m_{\alpha}\bar{\psi}_{\alpha}\psi_{\alpha} +\sum_{i}m_{i}\bar{\psi}_{i}\psi_{i}),  \nonumber \\
J_{\mu} & \sim &\sum_{\alpha,i}\bar{\psi}_{\alpha}V_{\alpha i}\psi_{i}.
\end{eqnarray}
Here, to highlight the part of the lepton-Higgs $\mathcal{L}$ which is relevant for our discussion,
we omit the gauge coupling constants and proper Dirac matrices in $\mathcal{L}(l,H)$.  Also, $\alpha=(e,\mu,\tau)$,
$\psi_{i}$ refers to $\nu_{i}$, $m_{\alpha}$ and $m_{i}$ are their masses, 
$W_{\mu}$ refers to the $W$-boson,  
$h$ denotes the Higgs field and $v$ is its VEV,  and $V_{\alpha i}$ is an element
of the PMNS matrix.

We may now study the action of g-permutation on $\mathcal{L}(l,H)$.
If the permutation only acts on the fermions (with $X_{ij}=X_{ij}^{(0)}$, $X_{\alpha \beta}=X_{\alpha \beta}^{(0)}$),
\begin{eqnarray}
X_{ij}^{(0)}:\psi_{i} \leftrightarrow \psi_{j}, \hspace{0.1in} (\psi_{k} \leftrightarrow \psi_{k}), \nonumber \\
X_{\alpha \beta}^{(0)}:\psi_{\alpha} \leftrightarrow \psi_{\beta}, \hspace{0.1in} (\psi_{\gamma} \leftrightarrow \psi_{\gamma}),
\end{eqnarray}
then clearly $\mathcal{L}(l,H)$ is not invariant and the g-permutation symmetry is broken.
However, we may include $(m_{\alpha},m_{i},V_{\alpha i})$ as dynamical variables which also transform under the action of 
$X_{ij}$ and $X_{\alpha \beta}$,  The structure of $\mathcal{L}(l,H)$ suggests that they transform like P-tensor.
So now we have
\begin{eqnarray}\label{psimv} 
X_{ij}:\psi_{i} \leftrightarrow \psi_{j}; \hspace{0.1in} m_{i} \leftrightarrow m_{j}; \hspace{0.1in}
V_{\alpha i} \leftrightarrow V_{\alpha j}, \nonumber \\
X_{\alpha \beta}:\psi_{\alpha} \leftrightarrow \psi_{\beta}; \hspace{0.1in} m_{\alpha} \leftrightarrow m_{\beta}; \hspace{0.1in}
V_{\alpha i} \leftrightarrow V_{\beta i}.
\end{eqnarray}
With these assignments and referring to the tensor analysis in Sec. II, it is evident that $\mathcal{L}(l,H)$ is invariant:
\begin{equation}\label{eq12}
(X_{ij},X_{\alpha \beta}):\mathcal{L}(l,H) \leftrightarrow \mathcal{L}(l,H).
\end{equation}
The symmetry group here is $S_{3}(l) \times S_{3} (\nu)$.  Exactly the same argument can be given for the quark sector, 
with the replacement of $(e,\mu,\tau)$ by $(u,c,t)$ and $(\nu_{1},\nu_{2},\nu_{3})$ by $(d,s,b)$.
We conclude that the SM has a g-permutation symmetry group, given by 
$S_{3}(u) \times S_{3}(d) \times S_{3}(l) \times S_{3}(\nu)=[S_{3}]^{4}$, that operate not only
on the fundamental fermions, but also on the masses and mixing parameters, which transform like P-vectors,
as indicated by the indices they carry.  

A more concrete interpretation of Eq. (\ref{eq12}) is to regard $\mathcal{L}(l,H)$ as an effective Lagrangian.
It is used as a starting point for calculations in flavour physics.  The results thus obtained  are expressed
in terms of the physical parameters contained in $\mathcal{L}(l,H)$.  Eq. (\ref{eq12}) then implies that these results must 
exhibit the $[S_{3}]^{4}$ symmetry. Some examples are cited in Sec. V.


The existence of the symmetry group $[S_{3}]^{4}$ can also be established from
another approach.  The diagonalization of the Yukawa coupling 
$(\sim \bar{\psi}_{L}Y\psi_{R}H)$,
\begin{equation}
Y=u_{L}^{\dag}Y_{D}u_{R},
\end{equation}
and the absorption of $(u_{L},u_{R})$ into the wave functions yield the Lagrangian in the mass eigenstate basis.
This procedure also constrains additional $U(3)$ transformations on $\psi$,
so that the global symmetry $(U(3) \times U(3) \times ....)$ of the gauge interaction Lagrangian
is broken down to $U_{B}(1) \times U_{L}(1)$,
as stated in the literature.  However, the solution to the diagonalization of a $3 \times 3$ matrix has a 
six $(3!)$-fold symmetry,
\begin{equation}
Y_{D}=X^{\dag}Y'_{D}X,
\end{equation}
where $X$ is a $3\times 3$ permutation matrix, e.g., 
\begin{equation}\label{}
X_{12}=
\left(\begin{array}{ccc}
   0& 1 & 0 \\
   1 & 0 & 0 \\
   0& 0 & 1 \\
    \end{array}
    \right). 
\end{equation}
The replacement of $(u_{L},u_{R})$ by $(Xu_{L},Xu_{R})$ corresponds to
an operation of $S_{3}$, which survives the diagonalization process, and is a
symmetry of the Lagrangian. Thus, the SM (with Dirac neutrinos) is found
to have the global symmetry group $[S_{3}]^{4} \times U_{B}(1) \times U_{L}(1)$.
The action of $[{S_{3}}]^{4}$ is given by Eq. (\ref{psimv}).


We now pause to consider the effect of rephasing invariance, which was glossed over earlier.  With rephasing,
the transformation of $V_{\alpha i}$ can acquire a phase:
\begin{equation}
X_{ij}:V_{\alpha i} \rightarrow \mbox{(phase)}\cdot V_{\alpha j}.
\end{equation}
This means that only rephasing invariant combinations of a set of $V_{\alpha i}$'s can have definite transformation
laws under exchange.  Two well-known combinations are $W_{\alpha i}=|V_{\alpha i}|^{2}$  
and the Jarlskog invariant \cite{Jarlskog:1985ht}, defined by  
\begin{equation}
\mbox{Im}(V_{\alpha i}V_{\beta j}V_{\alpha j}^{*}V_{\beta i}^{*})=J \sum_{\gamma k} e_{\alpha \beta \gamma}e_{ijk}.
\end{equation}
Thus, for physical variables, the transformation laws under exchange are (for the lepton sector):
\begin{eqnarray}\label{WJ} 
X_{ij}: W_{\alpha i} &\leftrightarrow& W_{\alpha j}; \hspace{0.1in} J \leftrightarrow -J, \nonumber \\
X_{\alpha \beta}: W_{\alpha i} &\leftrightarrow& W_{\beta i}; \hspace{0.1in} J \leftrightarrow -J.
\end{eqnarray}
Note that, in the terminology of Sec. II, $J$ is a pseudo-P-scalar.  Also, for the quark sector, 
the corresponding invariant, $J^{(q)}$, is also a pseudo-P-scalar:
\begin{equation}
(X_{ij}^{(q)},X_{\alpha \beta}^{(q)}):J^{(q)} \leftrightarrow - J^{(q)}.
\end{equation}
While the transformation of $W_{\alpha i}$ is not surprising, 
$J \rightarrow -J$ under any exchange is a remarkable property.

Another interesting aspect of rephasing invariance is that one can take out an overall phase from $V$
and demand, without loss of generality, that $\mbox{det}V=+1$, while restricting further rephasing by
$\mbox{det}P=+1$, where $P$ is a diagonal phase matrix \cite{Kuo:2005pf}.  Under this condition there  
are the following rephasing invariants:
\begin{equation}
\Gamma_{ijk}^{\alpha \beta \gamma}=V_{\alpha i}V_{\beta j}V_{\gamma k}=\mbox{Re}[\Gamma_{ijk}^{\alpha \beta \gamma}]-iJ,
\end{equation}
where $\alpha \neq \beta \neq \gamma$, $i \neq j \neq k$.  This yields an alternative definition for $J$:
\begin{equation}
J=-\mbox{Im}(V_{\alpha i}V_{\beta j}V_{\gamma k}), \hspace{0.1in} \mbox{det}V=+1.
\end{equation}

Since $\mbox{det}V$ changes sign under both $V \rightarrow -V$ and the exchange of rows or columns,
to keep $\mbox{det}V=+1$ under the exchange operation, we have now
\begin{eqnarray}\label{V} 
X_{ij}:V_{\alpha i} \leftrightarrow V_{\alpha j} \hspace{0.1in} \mbox{and} \hspace{0.1in} V \leftrightarrow -V, \nonumber \\
X_{\alpha \beta}:V_{\alpha i} \leftrightarrow V_{\beta i} \hspace{0.1in} \mbox{and} \hspace{0.1in} V \leftrightarrow -V.
\end{eqnarray}
Note that for $V \rightarrow -V$, the invariants $J$ and $W_{\alpha i}$ are unaffected.
The variables $(x_{i},y_{j})$ \cite{Kuo:2005pf}, which was defined in terms 
of $\mbox{Re}[\Gamma_{ijk}^{\alpha \beta \gamma}]$, now have the transformation laws:
\begin{equation}
(X_{ij} \hspace{0.1in} \mbox{or} \hspace{0.1in} X_{\alpha \beta}):(x_{1},x_{2},x_{3}) \leftrightarrow -(y_{a},y_{b},y_{c}),
\end{equation}
where $(a,b,c)$ is a permutation of $(1,2,3)$.  This implies, in particular, that $\sum x_{i} \leftrightarrow -\sum y_{j}$
and $x_{1}x_{2}x_{3} \leftrightarrow -y_{1}y_{2}y_{3}$.  With $J^{2}=x_{1}x_{2}x_{3}-y_{1}y_{2}y_{3}$, we have
$J^{2} \rightarrow J^{2}$, consistent with $J \leftrightarrow -J$.

In summary, the SM has the symmetry group $[S_{3}]^{4}$ when the physical mass matrix parameters behave as
tensors.  This set includes the fermion masses, the mixing matrices $[W^{(CKM)}]$ and $[W^{(PMNS)}]$,
and two signs for the Jarlskog invariants, $J=\pm \sqrt{J^{2}}$ ($J^{2}=\mbox{function of} \hspace{0.1in} [W]$), in the quark 
and lepton sectors, respectively.  The transformation laws are given in Eqs. (\ref{psimv}), (\ref{WJ}), and (\ref{V}).

\section{Composite Tensors}

To apply the g-permutation symmetry to physical processes, it turns out that, besides the basic tensors, certain
of their combinations make frequent appearances.  We now present a brief discussion of their properties.

A) For the masses $m_{\alpha}$ (and similarly for $m_{i}$), we have a scalar, $\sum m_{\alpha}$, and an anti-symmetric
tensor $\Delta m_{\beta \gamma}=m_{\beta}-m_{\gamma}$, which becomes a pseudo-P-vector:
\begin{equation}
\Delta \widetilde{m}_{\alpha}=\frac{1}{2}e_{\alpha \beta \gamma} \Delta m_{\beta \gamma} \sim \widetilde{\bf{3}}.
\end{equation}
This combination will appear repeatedly in applications.

B) Out of two $W_{\alpha i}$'s, we can have: 
\begin{itemize}
\item $W_{\alpha i}-W_{\alpha j}$, or $\frac{1}{2}e_{ijk}(W_{\alpha j}-W_{\alpha k})$, which transforms as a \textbf{3}
in $S_{3}(l)$ and a $\widetilde{\textbf{3}}$ in $S_{3}(\nu)$. 
\item $\frac{1}{2}e_{\alpha \beta \gamma}e_{ijk}W_{\beta j}W_{\gamma k}=w_{\alpha i}$. 
Here, $w_{\alpha i}$ is an element of the cofactor matrix of $W$, as defined before \cite{Kuo:2005pf}. 
Sums of its rows and columns are given by $\sum_{\alpha} w_{\alpha i}=\sum_{i} w_{\alpha i}=\mbox{det}W$.
It transforms as the product of pseudo-P-vectors $\widetilde{\bf{3}}(l) \times \widetilde{\bf{3}}(\nu)$, or
$\widetilde{\bf{3}}(u) \times \widetilde{\bf{3}}(d)$ in the quark sector.
For specific forms of the $[W]$ matrix, with a) $[W]=I$, and b)$[W]=[D_{0}]/3$ 
(maximal mixing, $[D_{0}]_{\alpha i}=1$, for all $\alpha$ and $i$), we have
a) $[w]=I$ and b) $[w]=0$.  These properties will be useful later. 
\item $\frac{1}{2}[\frac{1}{2}E_{\alpha \beta \gamma} E_{ijk} W_{\beta j} W_{\gamma k}-W_{\alpha i}]=\Lambda_{\alpha i}$.
This combination was also used before \cite{Chiu:2017ckv,Chiu:2016qra,Chiu:2015ega}. 
It played an essential role in many of the formulas in neutrino oscillation and in RGE.
The transformation properties of $\Lambda_{\alpha i}$ are exactly like $W_{\alpha i}$.  What sets them apart is
its structure for specific forms of the $[W]$ matrix.  Of particular interests are: i) if $W_{\alpha i}=0$, then $\Lambda_{\beta j}=0$,
$\alpha \neq \beta$, $i\neq j$, and ii) if $W_{\alpha i}=1$, then except possibly for $\Lambda_{\alpha i}$, all other
$\Lambda_{\beta j}=0$.  
iii) If $[W]=[D_{0}]/3$, for maximal mixing, then $[\Lambda]=-[D_{0}]/18$.
If $[W]=[I]$, then $[\Lambda]=[0]$.
To prove i), note that $W_{\alpha i}=0$ implies $V_{\alpha i}=0$.
One can then use the alternative definition $\Lambda_{\beta j}=\mbox{Re}[V_{\gamma k}V_{\delta l}V_{\gamma l}^{*}V_{\delta k}^{*}]$
to deduce $\Lambda_{\beta j}=0$.  As for ii), if $W_{\alpha i}=1$, then $W_{\alpha j}=W_{\beta i}=0$, $\alpha \neq \beta$, $i \neq j$.
Using i), ii) follows.  
\item $W_{\alpha j}W_{\alpha k}$, or $E_{ijk}W_{\alpha j}W_{\alpha k}$.
This is yet another composite which transforms
like $W_{\alpha i}$.  It is, however, not independent because of the relation:
\begin{equation}\label{LambdaW}
\Lambda_{\beta i}+\Lambda_{\gamma i}=-W_{\alpha j}W_{\alpha k}.
\end{equation}
Nevertheless, it is sometimes used for simplicity. 
\end{itemize}

C) We can also construct 
\begin{equation}
\frac{1}{3!}e_{\alpha \beta \gamma}e_{ijk}W_{\alpha i}W_{\beta j}W_{\gamma k}=\mbox{det}W.
\end{equation}
Under $X_{\alpha \beta}$ or $X_{ij}$, $\mbox{det}W \leftrightarrow -\mbox{det}W$, so that $\mbox{det}W$
is a pseudo-P-scalar.  So far, no practical use has been found for $\mbox{det}W$, 
nor for other higher rank tensors out of the basic ones.

\section{Applications}

We may now turn to detailed analyses of neutrino oscillations and RGE, in which one can
arrange to vary certain parameters to induce changes to all the parameters as a set.
The resulting equations will now be written in the tensor notation.
This new look offers fresh insights into their structure, making them more understandable.
In the following we will study these issues case by case.

\subsection{Neutrino Oscillation in Vacuum}

When a neutrino beam travels down a path, the neutrino mass eigenstates pick up a phase,
$\mbox{exp}(2i\phi_{i})$, $\phi_{i}=m^{2}_{i}L/4E$, which then causes change of the mixing matrix.
This effect depends only on the phase difference, $\Delta \phi_{ij}=\phi_{i}-\phi_{j}$.
In the tensor terminology, neutrino oscillation is driven by the pseudo-P-vector
\begin{equation}
\widetilde{\Phi}_{i}=\frac{1}{2!}e_{ijk}\Delta \phi_{jk}=\Delta \widetilde{m}_{i}^{2}(\frac{L}{4E}) \sim \widetilde{\bf{3}}.
\end{equation}
The probability $P(\nu_{\alpha} \rightarrow \nu_{\beta})$ is well known, and, in a notation
that is adoptable for our use, given by Eqs. (58) and (59) of Ref. \cite{Chiu:2017ckv}, 
\begin{equation}\label{aa}
P(\nu_{\alpha} \rightarrow \nu_{\alpha})=1-4(W_{\alpha 1}W_{\alpha 2}\sin^{2}\Phi_{21}+W_{\alpha 1}W_{\alpha 3}
\sin^{2}\Phi_{31}+W_{\alpha 2}W_{\alpha 3}\sin^{2}\Phi_{32}),
\end{equation}
\begin{eqnarray}\label{ab}
P(\nu_{\alpha}\rightarrow \nu_{\beta})=&-&4[\Lambda_{\gamma 3}\sin^{2}\Phi_{21} 
+ \Lambda_{\gamma 2}\sin^{2}\Phi_{31} 
+ \Lambda_{\gamma 1}\sin^{2}\Phi_{32}] \nonumber \\
&+& 2J[\sin 2\Phi_{21}+\sin 2\Phi_{13} + \sin 2\Phi_{32}],  \hspace{0.2in} \alpha \neq \beta \neq \gamma.
\end{eqnarray}
To transcribe these equations, we start with $P(\nu_{\alpha} \rightarrow \nu_{\beta})$, $\alpha \neq \beta$.
In tensor notation it reads
\begin{equation}
P(\nu_{\alpha} \rightarrow \nu_{\beta})=-4E_{\alpha \beta \gamma}\delta_{ij}\Lambda_{\gamma i}
\sin^{2}\widetilde{\Phi}_{j}+2J(\sum_{i}\sin 2\widetilde{\Phi}_{i}).
\end{equation}
Thus, first of all, $P(\nu_{\alpha}\rightarrow \nu_{\beta})$ is a P-scalar under $S_{3}(\nu)$, 
which is reasonable. This is achieved by combining $\Lambda_{\alpha i}$ with $\sin^{2}\widetilde{\Phi}_{i} (\sim \bf{3})$
and $J (\sim \widetilde{\bf{1}})$ with $\sum_{i}\sin 2\widetilde{\Phi}_{i}$ (also $\sim \widetilde{\bf{1}}$).
Note also that $P(\nu_{\alpha} \rightarrow \nu_{\beta})=0$ if $[W]=[I]$, and $[\Lambda]$ is
the unique matrix (not $[W]$ or $[w]$) which also vanishes if $[W]=[I]$.
The formula for $P(\nu_{\alpha} \rightarrow \nu_{\beta})$ shows that it consists of a symmetric part,
$\sim S_{\alpha \beta}$, and an antisymmetric part, $\sim J \sim \widetilde{\bf{1}}$.
The antisymmetric part is CP and T violating.  With $J$ being a pseudo-P-scalar, we can apply
$X_{\alpha \alpha'}$ repeatedly and obtain
\begin{equation}
P(\nu_{\alpha} \rightarrow \nu_{\beta})=P(\nu_{\beta} \rightarrow \nu_{\gamma})=P(\nu_{\gamma} \rightarrow \nu_{\alpha})
=-P(\nu_{\beta} \rightarrow \nu_{\alpha})= .....,
\end{equation}
which is a well-known result.  It should also be noted that, in the quark sector, the CP-measure,
$J \cdot \Pi \Delta m^{2}_{\alpha \beta} \cdot  \Pi \Delta m^{2}_{ij}$, is a P-scalar under $S_{3}(u) \times S_{3}(d)$,
again a reasonable requirement for general CP violations.

Finally, the probability $P(\nu_{\alpha}\rightarrow \nu_{\alpha})$ can be obtained by unitarity,
$\sum_{\beta} P(\nu_{\alpha}\rightarrow \nu_{\beta})=1$, with use of the relation in Eq.~(\ref{LambdaW}).

\subsection{Neutrino Oscillation in Matter}

When neutrinos propagate in a medium rich in electrons, the effective Hamiltonian acquires an induced
mass $(\delta H)_{ee}=A$.  We may regard this as the first component of a P-vector,
$(\delta H^{D}_{ee}, \delta H^{D}_{\mu \mu},\delta H^{D}_{\tau \tau})=(\delta H^{D})_{\xi} \sim \bf{3}$, 
which also covers the possibility of ``gedanken media" that are rich in $\mu$ and/or in $\tau$.
The addition of $(\delta H^{D})_{\xi}$ generates changes in the physical variables.
These were expressed \cite{Chiu:2017ckv} as differential equations given by,  
with $dA=(\delta H^{D})_{ee}$ and $D_{i}=m^{2}_{i}$,      

\begin{equation}\label{dDA}
\frac{dD_{i}}{dA}=W_{ei}, 
\end{equation} 
\begin{eqnarray}\label{dWdA}
\frac{1}{2} \frac{d}{dA}
\left(\begin{array}{ccc}
   W_{e1}& W_{e2} & W_{e3} \\
   W_{\mu1} & W_{\mu2} & W_{\mu3} \\
   W_{\tau1} & W_{\tau2} & W_{\tau3} \\
    \end{array}
    \right)
&=&\frac{1}{D_{1}-D_{2}}
\left(\begin{array}{ccc}
   W_{e1}W_{e2}, & -W_{e1}W_{e2},& 0\\
  \Lambda_{\tau3}, & -\Lambda_{\tau3}, & 0 \\
   \Lambda_{\mu3}, & -\Lambda_{\mu3}, & 0 \\
    \end{array}
    \right)  \nonumber \\ 
    &+&  \frac{1}{D_{2}-D_{3}}
    \left(\begin{array}{ccc}
   0, & W_{e2}W_{e3}, & -W_{e2}W_{e3} \\
   0, & \Lambda_{\tau1}, &-\Lambda_{\tau1} \\
    0, & \Lambda_{\mu1}, &-\Lambda_{\mu1} \\
    \end{array}
    \right)  \nonumber \\
& + & 
\frac{1}{D_{3}-D_{1}}
\left(\begin{array}{ccc}
   -W_{e1}W_{e3}, & 0, & W_{e1}W_{e3} \\
   -\Lambda_{\tau2}, & 0, & \Lambda_{\tau2} \\
    -\Lambda_{\mu2}, & 0, & \Lambda_{\mu2} \\
    \end{array}
    \right),
\end{eqnarray}   
\begin{equation}
\frac{d}{dA}(\ln J)=-\sum_{i\neq j}\frac{W_{ei}-W_{ej}}{D_{i}-D_{j}}.
\end{equation}

In tensor notation, these equation read
\begin{equation}\label{tm}
\delta m^{2}_{i}=(\delta H^{D})_{\xi} W_{\xi i},
\end{equation}
\begin{equation}\label{tW}
\delta W_{\alpha i}=2E_{\alpha \xi \eta}e_{ijk}(\delta H^{D})_{\xi} \Lambda_{\eta j}/\Delta \widetilde{D}_{k},
\end{equation}
\begin{equation}\label{tJ}
\delta (\ln J)=-(\delta H^{D})_{\xi}\Delta \widetilde{W}_{\xi k}/\Delta \widetilde{D}_{k}.
\end{equation}
Here, $\Delta \widetilde{D}_{k}=(\frac{1}{2})e_{klm}(D_{l}-D_{m})$, 
$\Delta \widetilde{W}_{\xi k}=(\frac{1}{2})e_{klm}(W_{\xi l}-W_{\xi m})$.
For normal medium we will take $(\delta H^{D})_{\xi}=(dA,0,0)$. In this case, Eq.~(\ref{tW}) does not
cover $\delta W_{ei}$, which can be obtained by 
\begin{equation}
\delta W_{ei}=-\delta (W_{\mu i}+W_{\tau i}).
\end{equation}

We can now discuss the salient features of these equations.  To begin with, it is clear that
they are covariant tensor (including P-parity) equations under $S_{3}(l) \times S_{3}(\nu)$.
It is noteworthy that these equations utilize the tensors $W_{\alpha i}$, $\Lambda_{\alpha i}$, and $\Delta \widetilde{W}_{\alpha i}$
(but not $w_{\alpha i}$ or $E_{ijk}W_{\alpha j}W_{\alpha k}$), 
all of which have similar transformation properties.
It turns out that there are consistency conditions which dictate where they belong.
For Eq.~(\ref{tm}), in the special case $[W]=[I]$, it is known that $\delta m^{2}_{1}=(\delta H^{D})_{ee}$,
thus ruling out the use of $\Lambda_{\xi i}$, since $[\Lambda ]=[0]$ for $[W]=[I]$.
For Eq.~(\ref{tW}), we note that since $0 \leq W_{\alpha i} \leq 1$, it is necessary that $\delta W_{\alpha i}=0$
at the boundary $W_{\alpha i}=0$ or $1$.  From the properties listed in Sec. II, we find that  
$\Lambda_{\eta j}=0$ if $W_{\alpha i}=0$ or $1$, for $\eta \neq \alpha$, $j \neq i$.
These conditions are exactly met owing to the factor $E_{\alpha \xi \eta}e_{ijk}$, so that
from Eq.~(\ref{tW}), $\delta W_{\alpha i}=0$ if $W_{\alpha i}=0$ or $1$.  
Finally, for Eq.~(\ref{tJ}), there are also two consistency checks.
First, $J^{2}$ is known
to have a maximum at $[W]=[D_{0}]/3$, with $J^{2}_{max} = 1/108$.
Also, $J<0$ and $J>0$ belong to two separate regimes reachable by discrete transformations,
but not by infinitesimal increments.  It follows that we must demand that $\delta J=0$
for $|J|=J_{max}$ or $J=0$.  This also means that $\delta J=0$ if any $W_{\alpha i}=0$ or $1$, since
these last conditions imply $J=0$.  To satisfy the requirement at $J_{max}$, $\Delta \widetilde{W}_{\xi k}$ is
a possible (with wrong parity?) candidate in Eq.~(\ref{tJ}).  But the second condition, 
that $\delta J=0$ if any $W_{\alpha i}=0$ or $1$, can not be fulfilled by any tensor.
Eq.~(\ref{tJ}) solves this problem (and the P-parity problem) by using $\ln J$ so that
$\delta J \sim J[......]$, and $\delta J=0$ if $J=0$.  It is remarkable how the symmetry argument
and the direct calculations reinforce each other in confirming these equations.

\subsection{RGE for Quarks and Neutrinos}

In this section, we will deal exclusively with the RGE for quarks.  The RGE for Dirac neutrinos are
almost identical (see Eqs. (26) and (33) in Ref. \cite{Chiu:2015ega}), 
but are simpler since terms proportional to neutrino masses can be dropped.
It is thus sufficient to consider quarks only.

One loop RGE for quark mass matrices have been studied for a long time.  When written in the
matrix form, they are given by \cite{Machacek:1983fi,Sasaki:1986jv,Babu:1987im} 
\begin{equation}\label{6A}
\mathcal{D} M_{u}=a_{u}M_{u}+bM^{2}_{u}+c\{M_{u},M_{d}\},
\end{equation}
\begin{equation}\label{6B}
\mathcal{D} M_{d}=a_{d}M_{d}+bM^{2}_{d}+c\{M_{u},M_{d}\}.
\end{equation}
Here, $M_{u}=Y_{u}Y_{u}^{\dag}$ ($M_{d}=Y_{d}Y_{d}^{\dag}$), $Y_{u}$ ($Y_{d}$) is the
Yukawa coupling matrix for the $u$-type ($d$-type) quarks; $\mathcal{D}=(\frac{1}{16 \pi^{2}})\frac{d}{dt}$,
$t=\ln(\mu/M_{\bf{W}})$, $\mu$ is the energy scale, and $M_{\bf{W}}$ is the $\bf{W}$ boson mass.
The model dependence of the RGE is contained in the constants ($a_{u},a_{d},b,c$).

Although the RGE for the mass matrices are simple, they are not directly useful since the matrices contain
a large number of unphysical degrees of freedom.  One needs to extract the RGE for the physical parameters.
This was carried out, but usually in variables which mask the underlying symmetry.
The easiest for our  adaptation are the equations obtained in Ref. \cite{Chiu:2008ye,Chiu:2016qra}. 
These equations describe the variations of the mass ratios, the mixing parameters, and $J$.

We now write down the tensor form of these equations, and then justify them by comparing with the
established ones which were obtained by direct and explicit calculations.  In the following, as before,
the indices $(\alpha, \beta, \gamma)$ and $(i,j,k)$ refer to $(u,c,t)$ and $(d,s,b)$, respectively.
We define the mass ratios, 
\begin{equation}
R_{\alpha \beta}=\frac{m_{\alpha}^{2}}{m^{2}_{\beta}}, \hspace{0.1in} \ln\widetilde{R}_{\alpha}=
\frac{1}{2}e_{\alpha \beta \gamma}\ln R_{\beta \gamma},
\end{equation}
\begin{equation}
r_{ij}=\frac{m_{i}^{2}}{m^{2}_{j}}, \hspace{0.1in} \ln\widetilde{r}_{i}=
\frac{1}{2}e_{ijk}\ln r_{jk}.
\end{equation}
Also, the mass differences,
\begin{equation}
\Delta \widetilde{m}_{\alpha}^{2}=\frac{1}{2}e_{\alpha \beta \gamma}(m_{\beta}^{2}-m_{\gamma}^{2}),
\end{equation}
\begin{equation}
\Delta \widetilde{m}_{i}^{2}=\frac{1}{2}e_{ijk}(m_{j}^{2}-m_{k}^{2}).
\end{equation}
Finally, the combinations,
\begin{equation}
\widetilde{H}_{\alpha}=e_{\alpha \beta \gamma}H_{\beta \gamma}, \hspace{0.1in}
H_{\beta \gamma}=\frac{m_{\beta}^{2}+m_{\gamma}^{2}}{ m^{2}_{\beta}-m^{2}_{\gamma}},
\end{equation}
\begin{equation}
\widetilde{G}_{i}=e_{ijk}G_{jk},  \hspace{0.1in}
G_{jk}=\frac{m_{j}^{2}+m_{k}^{2}}{m^{2}_{j}-m^{2}_{k}}.
\end{equation}
Note that they transform as tensors according to the indices they carry, including pseudo-P-tensors
which are identified  with a ``$\sim$" symbol.

With these we can write down the following RGE in tensor form,
\begin{equation}\label{8A}
\mathcal{D}\ln \widetilde{R}_{\alpha}=b' \Delta \widetilde{m}_{\alpha}^{2}+2c' \cdot w_{\alpha i}\Delta \widetilde{m}_{i}^{2},
\end{equation}
\begin{equation}\label{8B}
\mathcal{D}\ln \widetilde{r}_{i}=b' \Delta \widetilde{m}_{i}^{2}+2c' \cdot w_{i \alpha }^{T}\Delta \widetilde{m}_{\alpha}^{2},
\end{equation}

\begin{equation}\label{8C}
\mathcal{D}W_{\alpha i}=-2c' \cdot (\Delta \widetilde{m}^{2}_{\beta}[S^{\alpha i}]_{\beta j}\widetilde{G}_{j}+
\Delta \widetilde{m}^{2}_{j}[S^{\alpha i}]^{T}_{j\beta}\widetilde{H}_{\beta})
\end{equation}

\begin{equation}\label{8D}
\mathcal{D}\ln J=-c' \cdot (\Delta \widetilde{m}^{2}_{\alpha}w_{\alpha i}
\widetilde{G}_{i}+\Delta \widetilde{m}_{i}^{2}w^{T}_{i \alpha}\widetilde{H}_{\alpha}).
\end{equation}
In Eq.(49), the matrix elements of $[S^{AI}]$ are given by $[S^{AI}]_{BJ}=(\sum_{\gamma k}e^{AB\gamma}e^{IJk})\Lambda_{BJ}$.
These equations are just the equations (3.17) and (3.18) in Ref.\cite{Chiu:2008ye}   
and equations (28) and (29) in Ref. \cite{Chiu:2016qra}, although the indices $\alpha$ and $i$ were not
distinguished, nor were the tensors clearly identified.
Also, $b'=b/v^{2}$, $c'=c/v^{2}$, since masses are used directly here.

The first thing that catches the eye in these equations is that they are manifestly covariant tensor
equations under $S_{3}(u) \times S_{3}(d)$.  Let us now concentrate on the $c$-dependent part of them.
The resemblance to Eqs.~(\ref{tm}), (\ref{tW}), and (\ref{tJ}) is striking, although there are also differences.  For
Eqs.~(\ref{8A}), (\ref{8B}), (\ref{8C}), and (\ref{8D}), there are two ``source" terms, 
$\Delta \widetilde{m}_{i}^{2}$ and $\Delta \widetilde{m}_{\alpha}^{2}$,
which generate the changes.  Their effects on $V_{u}^{\dag}$ and $V_{d}$ are combined in $V_{CKM}$,
and thus in $\mathcal{D}W_{\alpha i}$ and $\mathcal{D}(\ln J)$.  Also, the simple pole terms
$[1/(D_{i}-D_{j})]$ in Eqs.~(\ref{tW}) and (\ref{tJ}) are replaced by $G_{ij}$ and $H_{ij}$, reflecting the nature of the  
new situation, while keeping the singular behaviour if $(D_{i}-D_{j}) \rightarrow 0$.  In addition,
$(\delta H^{D})_{\xi} \sim \bf{3}$ for neutrino oscillations, while $\Delta \widetilde{m}^{2}_{\alpha}$ 
and $\Delta \widetilde{m}^{2}_{i} \sim \widetilde{\bf{3}}$ for RGE.  
Note that $[S^{\alpha i}]$ satisfies the consistency conditions $\sum_{\alpha}[S^{\alpha i}]=\sum_{i}[S^{\alpha i}]=[0]$
and, if $W_{\alpha i}=0$ or 1, $[S^{\alpha i}]=[0]$, ensuring that $\mathcal{D}W_{\alpha i}=0$ when $W_{\alpha i}$
assumes its extremal values. Also, the tensor $[S^{\alpha i}]_{\beta j}$ transforms as $\bf{3} \times \bf{3}$
and  $\widetilde{\bf{3}} \times  \widetilde{\bf{3}}$ with respect to its upper and lower indices, respectively.   
As for $\mathcal{D} (\ln J)$, P-parity calls for switching $\Delta \widetilde{W}_{\xi k}$ to $w_{\alpha i}$, and all
consistency requirements are met.  Now a comment on terms that depend on $a$ or $b$ in Eqs.~(\ref{6A}) and (\ref{6B}).
These terms do not contribute to changes in mixing, since the diagonalization of $M$ is the same as that of a polynomial
in $M$.  This is also why only mass differences, $\Delta \widetilde{m}_{\alpha}^{2}$ and $\Delta \widetilde{m}_{i}^{2}$,
appear in these equations.  A common mass in $m_{\alpha}^{2}$ or $m_{i}^{2}$, according to Eqs.~(\ref{6A}) and (\ref{6B}),
can always be absorbed in $a_{d}$ and $a_{u}$, respectively.

Just as for neutrino oscillations, it is impressive to see how the results of direct calculations fit
into the framework of permutation symmetry.  Conversely, except for some overall constants, 
and barring the use of higher-rank tensors [e.g., $(\mbox{det}W)^{2}\Lambda_{\alpha i}$],
one could almost write down these equations without any detailed computations.


Another consequence of permutation symmetry is that tensors are the entities being measured.
E.g., neutrino oscillations determine the tensors $\Lambda_{\gamma k}$.
They, in turn, are simple functions of $W_{\alpha i}$. It is therefore useful to analyze 
data directly in terms of $W_{\alpha i}$, thereby avoiding the possible loss of information
in translation. Some of the issues were also discussed elsewhere \cite{Chiu:2017ckv}.

\section{conclusion}
 
The SM is notorious for having a multitude of parameters.  They originate from the breaking
of the g-permutation symmetry by the Higgs interaction.  In this paper we suggest that, instead of
regarding them as fixed numbers, these parameters can be included as physical variables which also transform under the actions
of the g-permutation operation.  By assigning them as appropriate tensors, the symmetry is shown to be
restored.  Indeed, using this procedure, the SM (with inclusion of Dirac neutrino mass terms)
is found to have the discrete symmetry,
$S_{3}(u) \times S_{3}(d) \times S_{3}(l) \times S_{3}(\nu)=[S_{3}]^{4}$.

We apply the symmetry to physical processes, including neutrino oscillations and RGE, for which
there are established relations between the parameters due to a change in certain variables.
These relations are the results of direct calculations.  When we rewrite them in terms of tensors,
it is revealed that they are indeed covariant tensor equations under $[S_{3}^{4}]$.  The tensor
notation helps to make them very compact and simple in form.  In addition, one gains insights that could otherwise
be obscured by a different notation.  It is hoped that there will be further developments and applications 
of the symmetry principle in other areas of flavor physics.

\acknowledgments                 
SHC is supported by the Ministry of Science and Technology of Taiwan, 
Grant No.: MOST 107-2119-M-182-002.



\end{document}